\documentclass[a4paper,12pt,english,german]{article}
\usepackage{graphicx}

\begin{document}

\centerline{\bf Parallel decoherence in  composite quantum
systems}

\bigskip

\centerline{M. Dugi\' c$^a$, J. Jekni\' c-Dugi\' c$^b$}

\bigskip

\centerline{$^a$Department of Physics, Faculty of Science,
Kragujevac, Serbia}

\centerline{$^b$Department of Physics, Faculty of Science and
Mathematics, Ni\v s, Serbia}

\bigskip

{\bf Abstract} For the standard Quantum Brownian Motion (QBM)
model, we point out the occurrence of  simultaneous (parallel),
mutually irreducible and autonomous decoherence processes. Besides
the standard, one Brownian particle, we show there is at least
another one system undergoing the dynamics described by the QBM
effect. We do this by selecting the two mutually irreducible,
global structures (decompositions into subsystems) of the
composite system of the QBM model. A generalization of this
observation is a new, challenging task in the foundations of the
decoherence theory. We do not place our findings in any
interpretational context.

\bigskip


\bigskip

PACS 03.65.Yz {Decoherence; open systems; quantum statistical
methods}

PACS 03.65.Ud {Entanglement and quantum nonlocality (e.g. EPR
paradox, Bell's inequalities, GHZ states, etc.)

PACS 03.65.Ta {Foundations of quantum mechanics; measurement
theory}

\bigskip

{\bf 1. Introduction}

\bigskip

\noindent "{\it In particular, one issue which has been often
taken for granted is looming big, as a foundation of the whole
decoherence program. It is the question of what are the 'systems'
which play such a crucial role in all the discussions of the
emergent classicality. (. . . ) [A] compelling explanation of what
are the systems--how to define them given, say, the overall
Hamiltonian in some suitably large Hilbert space--would be
undoubtedly most useful.}" (p. 1820 in Ref. [1]).

In this paper, we consider the two specific structures
(decompositions, partitions into subsystems) of the standard
Quantum Brownian Motion (QBM) setup [2, 3, 4, 5] and we obtain the
QBM effect for both structures considered. The structures are
mutually irreducible (i.e. can {\it not} be obtained from each
other by decomposing, grouping or permutations of the constituent
subsystems) and global (do {\it not} have even a single degree of
freedom in common). The structures are mutually linked by the
linear canonical transformations (LCTs) thus being dynamically
independent, autonomous structures of the one and the same
composite system of the QBM model that cannot be  obtained from
each other via the "coarse graining" [6] operation.

The LCTs is a universal physical method. Already the unitary
evolution regroups or separates the constituent subsystems thus
locally changing a structure of a composite system. In quantum
decoherence, the LCTs are sometimes used to ease the calculation
[5], while (kinematically) regrouping or  decomposing the
subsystems may shed some new light on the mechanism of decoherence
[7, 8]. A change in the environmental degrees of freedom reveals
some subtleties such as the "system-size" dependence of
decoherence [5, 9] and can help in distinguishing the robust (the
preferred "pointer basis" [4, 5, 10, 11]) states for the open
system [12, 13, 14]. However, these models and considerations
refer to the {\it local} structures that share the degrees of
freedom or can be reduced to each other. This makes the structures
considered mutually dynamically coupled or dependent, which is not
our objective.

Paradigmatic for our considerations is the Hydrogen Atom (HA)
model. The quantum theory of the hydrogen atom relies on the
transformations of the electron's ($e$) and the proton's ($p$)
degrees of freedom to introduce the atom center of mass ($CM$) and
the "relative position ($R$)" degrees of freedom. Due to the
absence of the coupling between $CM$ and $R$, one obtains the
variables separation and the exact solution of the atomic internal
energy and eigenstates. The two structures of HA, $e+p$ and
$CM+R$, are mutually both {\it irreducible} and {\it global}
likewise those of the QBM setup we discuss below.

In Section 2, we derive our main result on the parallel
decoherence: at variance with the standard wisdom, we point out
that a composite quantum system can be described by (may host) the
different, simultaneously existing and mutually independent
quasi-classical (global) structures. In Section 3 we generalize
our considerations and we emphasize that the "parallel
decoherence" launches a new task in the foundations of the
decoherence theory. In general, this task can be formidable yet
possibly of a wider scientific interest.  Section 4 is Discussion,
where we emphasize: as long as one can rely on our model-dependent
finding, if the standard decoherence program provides the
"appearance of {\it a} Classical World",  our results suggest the
'appearance of {\it the} Classical World{\it s}'. Nevertheless, we
do not enter any interpretational details. Section 5 is
Conclusion.

\bigskip

{\bf 2. Parallel decoherence in the QBM setup}

\bigskip

\noindent {\it  "Note that decoherence derives from the
presupposition of the existence and the possibility of a division
of the world into 'system(s)' and 'environment'.}" (p. 83 in Ref.
[11]).

By "decoherence" we mean the {\it environment-induced selection}
of the preferred ("pointer basis") states of an open quantum
system [4, 5, 10, 11, 15]. The effect of decoherence refers to
certain (typically "collective") observables (formally subsystems)
of a larger (open) system $S$ in unavoidable interaction with its
environment $E$. The interaction in $S+E$ system and its strength
determine a preferred set of (not necessarily orthogonal) pointer
basis states of the open system $S$ that bear robustness--these
states exhibit the "least" respond to the environmental influence.

For the sufficiently large environment, the robustness of the
pointer basis states  gives rise to the quasi-diagonal form of the
system's state $\hat \rho_S$ (in the pointer basis representation)
and the effect of decoherence appears to be irreversible in
practice, Zeh in [4] and [16]. This apparent irreversibility is a
physical basis of both the approximate quasi-classical dynamics of
the decohered system (a subsystem of $S$) as well as of the
classical-information contents of the open system $S$.

This typical attitude of the decoherence theory in introducing a
composite system $\mathcal{C}$ by {\it grouping} the actual
systems, $S$ and $E$, ($\mathcal{C} = S+E$) is at  variance with
our concern here. Actually, we admit existence of a (practically
closed) composite system $\mathcal{C}$ whose structure appearing
through the alternative decompositions into subsystems is yet to
be defined. The different structures are mutually linked by some
LCTs and we investigate the occurrence of decoherence for some
alternative structures.

In some abstract terms, our task can be readily formulated:
whether or not, the LCTs can provide the occurrence of decoherence
for an alternative structure of a composite system? While this is
the subject of the next section, here we just emphasize: as the
occurrence of decoherence is directly related to the interaction
term in the Hamiltonian of the composite system [15], the LCTs
should preserve the desired characteristics [15] of the
interaction even for the alternative structure. In this section,
we are concerned with the concrete, QBM model of the composite
system.

In order to emphasize our basic observation, we first deal with
the simplified model of a pair of the one-dimensional systems $S,
E$ linearly interacting through their respective position
operators $\hat x_S$ and $\hat x_E$.

The Hamiltonian is given by:

\begin{equation}
\hat H = {\hat p_S^2 \over 2m_S} + {\hat p_E^2 \over 2 m_E} + {m_E
\omega^2 \over 2} \hat x_E^2 - C \hat x_S \hat x_E \equiv \hat H_S
+ \hat H_E + \hat H_{S+E}.
\end{equation}

The standard LCT that introduce the center-of-mass ($CM$) and the
"relative positions" ($R$) observables ($\hat X_{CM} = (m_S\hat
x_S + m_E\hat x_E)/(m_S + m_E)$ and $\hat \rho_S = \hat x_S - \hat
x_E$, respectively) give rise to:

\begin{eqnarray}
&\nonumber& \hat H = {\hat p_{CM}^2 \over 2(m_S + m_E)} + c_1 \hat
X_{CM}^2  + {\hat p_R^2 \over 2 \mu} + c_2 \hat \rho_R^2 \\&& -
c_3 \hat X_{CM} \hat \rho_R \equiv \hat H_{CM} + \hat H_R + \hat
H_{CM+R};
\end{eqnarray}

\noindent $c_1 = m_E \omega^2/2 - C$, $c_2 = m_S \mu
\omega^2/2(m_S + m_E) + C \mu/(m_S + m_E)$, $c_3 = C (m_E -
m_S)/(m_S + m_E) + \mu \omega^2$ and $\mu = m_S m_E/(m_S + m_E)$
is the reduced mass, with the constraint $C < m_E \omega^2 / 2$.
This way, the composite system $\mathcal{C} = S + E$ is formally
redefined to introduce the alternative structure defined by the
"new" subsystems $CM$ and $R$. Certainly, $S + E = \mathcal{C} =
CM + R$ and the Hamiltonian eqs. (1) and (2) is the composite
system's Hamiltonian, $\hat H \equiv \hat H_C$.

The two structures, $S+E$ and $CM+R$, are mutually irreducible
(can not be obtained from each other by decomposing  or grouping
the constituent subsystems) and "global" (not having the common
degrees of freedom). The models are formally similar: the {\it
interaction terms} are exactly of the {\it same form} that
distinguishes the position-eigenstates as the candidate pointer
basis states [15] for {\it both} $S$ and $CM$. For the
many-particles environment (cf. below), both the open systems can
be described by the following master equation [4, 5]:

\begin{equation}
\imath \hbar {d \hat \rho \over dt} = [\hat H, \hat \rho] - \imath
\Lambda [\hat x, [\hat x, \hat \rho]],
\end{equation}

\noindent which models the position-observable ($\hat x$)
measurement. Therefore, the models, albeit simplified, suggest the
possible simultaneous and mutually independent existence of the
preferred states for the open systems of {\it both} decompositions
regarded.

Let us extend the model eq. (1) by identifying $\hat x_E$ with the
collective variable $\sum_i \kappa_i \hat x_{Ei}$ for the harmonic
bath ($E$). Then the model eq. (1) resembles the Caldeira-Leggett
model [2] for the quantum Brownian motion (QBM). For this model,
the occurrence of decoherence is a well-established result [2, 3,
4, 5]. Physically, the bath $E$ acts as a "quantum apparatus"
measuring the $S$'s position-observable $\hat x_S$. Interestingly
enough, for the initial Gaussian states for $S$, the occurrence of
decoherence for the linear position-observables coupling (also for
eq. (1) and eq. (2)) is {\it largely independent} of the details,
such as the appearance/nonappearance and the kind of the external
field $V(\hat x_S)$ in the model, the strength of interaction, the
spectral density or the bath's temperature  as well as of the
presence of the (classical or quantum) correlations in the initial
state of the composite system [3, 5]. The Gaussian states (that
include the standard "coherent states") appear as the approximate
pointer basis.

Let us formally consider the Caldeira-Leggett model:

\begin{eqnarray}
&\nonumber& \hat H = {\hat p_S^2 \over 2m_S} + V(\hat x_S) +
\sum_i ({\hat p_{Ei}^2 \over 2 m_i} + {m_i \omega_i^2 \hat
x_{Ei}^2 \over 2}) \\&& \pm \hat x_S \sum_i \kappa_i \hat x_{Ei}
\equiv \hat H_S + \hat H_E + \hat H_{S+E},
\end{eqnarray}

\noindent where the index $i$ enumerates the environmental
"particles", and the sign $\pm$ is in accord with the variations
of the model in the literature. The Hamiltonian eq. (4) generates
the unitary dynamics for the initially separable state of the
composite system, $\hat \rho_C = \hat \rho_S \otimes \hat
\rho_{Eth}$, where $\hat \rho_{Eth}$ denotes the
thermal-equilibrium state of the environment $E$. So, we consider
the standard, linear QBM setup with the separable initial state of
the composite system $\mathcal{C}$.

Now, we apply the standard LCTs introducing the center-of-mass
($CM$) and the relative-position variables for the {\it whole}
composite system $\mathcal{C}$, where the set of the relative
positions is collectively denoted as the subsystem $R$, $\{\hat
\rho_{R\alpha}\}$. Then, the inverse transformations give $\hat
x_i = \hat X_{CM} + \sum_{\alpha} \omega_{\alpha i} \hat
\rho_{R\alpha}$, and $\hat x_1 \equiv \hat x_S$, $\omega_{1 i}
\equiv \omega_{Si}$; $\omega$s can be positive/negative real
constants.

For the 'new' structure $CM + R$ one obtains:

\begin{eqnarray}
&\nonumber& \hat H = {\hat P_{CM}^2 \over 2M} + {1 \over 2} M
\Omega_{CM}^2 \hat X_{CM}^2+ \sum_{\alpha} ({\hat p_{R\alpha}^2
\over 2 \mu_{\alpha}} + {1 \over 2} \mu_{\alpha} \nu_{\alpha}^2
\hat \rho_{R\alpha}^2)  \\&& + \hat V_R \pm \hat X_{CM}
\sum_{\alpha} \sigma_{\alpha} \hat \rho_{R\alpha}.
\end{eqnarray}

\noindent for the two relevant models, of the free particle and of
the harmonic oscillator as the open system $S$. Introducing the
total ($CM$) mass $M$, the standard reduced masses $\mu_{\alpha}$
and the "mass polarization" constants $C_{\alpha \alpha'} =
m_{\alpha + 1} m_{\alpha' + 1}/M$, the constants appearing in eq.
(5) are as follows:

i. for the {\it free particle} ($V(\hat x_{S}) = 0$): $M
\Omega_{CM}^2/2 = \sum_i (\pm \kappa_i + m_i \omega_i^2/2)$,
$\mu_{\alpha} \nu_{\alpha}^2 /2 = \pm \omega _{\alpha S} \sum_i
\kappa_i \omega_{\alpha i} + \sum_i m_i \omega_i^2 \omega_{\alpha
i}^2 / 2$, and $\sigma_{\alpha} = \sum_i (\kappa_i \omega_{\alpha
i} + \kappa_i\omega_{\alpha S} + m_i \omega_i^2 \omega_{\alpha
i})$. The internal interaction term $\hat V_R = \sum_{\alpha \neq
\alpha'} [C_{\alpha \alpha'} \hat p_{R\alpha} \hat
p_{R\alpha'}/\mu_{\alpha}\mu_{\alpha'} + (\Omega_{\alpha \alpha'}
+ \omega_{\alpha S} \Omega_{\alpha'}) \hat \rho_{R\alpha} \hat
\rho_{R\alpha'}]$; $\Omega_{\alpha} = \sum_i \kappa_i
\omega_{\alpha i}$ and $\Omega_{\alpha \alpha'} = \sum_i m_i
\omega_i^2 \omega_{\alpha i} \omega_{\alpha' i}/2$. The conditions
of positivity, $M \Omega_{CM}^2/2 > 0$ and $\mu_{\alpha}
\nu_{\alpha}^2 /2 > 0$, exhibit the subtleties concerning the
choice of the physically interesting LCT;

ii. for the {\it harmonic oscillator} ($V(\hat x_S) = m_S
\omega_S^2 \hat x_S^2/2$): the harmonic part for $S$ adds the
terms appearing by the virtue of the inverse transformation (cf.
above): $\hat x_S = \hat X_{CM} + \sum_{\alpha} \omega_{\alpha S}
\hat \rho_{R\alpha}$. Particularly, the harmonic term for $S$
obtains the form: $m_S \omega_S^2 \hat X_{CM}^2 /2 + \sum_{\alpha}
m_S \omega_S^2 \omega_{\alpha S}^2 \hat \rho_{R\alpha}^2 /2 +
\sum_{\alpha \neq \alpha'} m_S \omega_S^2 \omega_{\alpha S}
\omega_{\alpha' S} \hat \rho_{R\alpha} \hat \rho_{R \alpha'}/2 +
\hat X_{CM} \sum_{\alpha} m_S \omega_S^2 \omega_{\alpha S} \hat
\rho_{R\alpha}$. By adding this sum to the Hamiltonian for the
free particle (the above case i.) one obtains the Hamiiltonian of
the general form eq. (5).

To this end, it is essential to note: the two structures of the
composite system, $S+E$ and $CM+R$, do {\it not} follow from each
other via decomposing, grouping or the permutations of subsystems
(degrees of freedom) operations. As the two structures do not have
even a single degree of freedom in common, they are "global", as
distinct from the "local" structures emphasized in Introduction
[5-8, 11-13]. The two open systems, $S$ and $CM$ are both
one-dimensional systems and can not be decomposed--they do not
posses any structure of their own. So, the two structures are
mutually {\it irreducible} and their unitary (Schrodinger)
dynamics are mutually {\it independent} and {\it autonomous}.

Compare the two models, eq. (4) and eq. (5), that equally apply to
the both cases i. and ii. The simple exchange of $CM$ and $R$ in
eq. (5) by $S$ and $E$ gives a formal variation of eq. (4). Both
open systems ($S$ and $CM$) are one-dimensional. Then, by the
virtue of the LCTs, the respective environments ($E$ and $R$) bear
the same number of the degrees of freedom--the same complexity and
ability to provide the "genuine decoherence". Both environments
are the harmonic-oscillators systems. The interaction terms are of
the exactly the {\it same form} that provides the "spectral
densities" [5] for the two models are of the same form.

The differences come about as follows. First for the Hamiltonian
forms that differ in the values of the parameters (the masses and
the characteristic frequencies of the oscillators), there is a new
harmonic term for $CM$ system relative to the system $S$, and
there is the (small-norm) term $\hat V_R$ involving the  couplings
for the new-environment's ($R$'s) oscillators. Second, the
variables transformations typically induce a change in quantum
state of the composite system: if a state is separable for one
structure, it is typically entangled for the alternative
structure--the {\it entanglement relativity} [17-24].

Nevertheless, as we show in Appendix 1, all these distinctions do
not change the conclusion presented above for the simple model.
Actually, in Appendix 1 we show that the composite system's
Hamiltonian $\hat H$ generates for the fixed initial state $\hat
\rho_C$ the two, simultaneously (in parallel) occurring and
mutually irreducible and independent  decoherence processes for
the two open systems $S$, and $CM$. While the details regarding
the occurrence of decoherence (such as the decoherence time, the
recurrence time or the state fluctuations) may be different for
the different structures, one can say: likewise the open system
$S$, the open system $CM$ is a 'Brownian particle' for its
respective structure. As the two decoherence processes unfold
simultaneously (in parallel) and are mutually irreducible and
independent (autonomous), we emphasize our main result as follows:
the isolated composite system $\mathcal{C}$ {\it hosts} at least
the two {\it simultaneously and mutually independently occurring}
(the parallel) decoherence processes that amount to the
approximate quasi-classical behavior of the  subsystems, i.e. of
the two, mutually irreducible and dynamically autonomous 'Brownian
particles', $S$ and $CM$.

\bigskip

{\bf 3. Some general considerations}

\bigskip

\noindent Let us present our considerations in the more abstract
terms.

The LCTs can be formally presented in a compact form as:

\begin{eqnarray}
 &\nonumber&\hat X_{S'\alpha} = X (\hat x_{Si}, \hat p_{Si}; \hat \xi_{Ej}, \hat \pi_{Ej}),
 \hat P_{S'\alpha} = P (\hat x_{Si}, \hat p_{Si}; \hat \xi_{Ej}, \hat \pi_{Ej}) \nonumber\\&&
 \hat \Xi_{E'\beta} = \Xi (\hat x_{Si}, \hat p_{Si}; \hat \xi_{Ej}, \hat \pi_{Ej}) ,
\hat \Pi_{E'\beta} = \Pi (\hat x_{Si}, \hat p_{Si}; \hat \xi_{Ej},
\hat \pi_{Ej}).
 \end{eqnarray}

\noindent
 In eq. (6) appear the (continuous) position- and the momentum- observables of the subsystems indicated as the indices to
 the respective observables. The LCTs do not assume any constraints on the degrees of freedom and the number of the degrees
 of freedom is conserved; the tensor-product structures of the $\mathcal{C}$'s Hilbert state fulfill the
 equality $\otimes_{i=1}^{\nu_S} H_{Si} \otimes_{j=1}^{\nu_E} H_{Ej} = \otimes_{p=1}^{\nu_{S'}} H_{S'p} \otimes_{q=1}^{\nu_{E'}} H_{E'q}$,
 while $\nu_{S} + \nu_E = \nu_{S'} + \nu_{E'}$. The LCTs are {\it global} if the related structures, $\mathcal{S} = \{\hat x_{Si}, \hat \xi_{Ej}\}$
 and $\mathcal{S'} = \{\hat X_{S'\alpha},
 \hat \Xi_{E'\beta}\}$, do not
 have a single degree of freedom in common, $\mathcal{S} \bigcap \mathcal{S'} =
 0$, and  the two structures are both subject to the Schr\"odinger
 law and are dynamically mutually independent. Further on,
 likewise in Section 2, we consider the mutually irreducible
 structures.

In the general terms, our task refers to a pair of (global)
decompositions, $S+E$ and $S'+E'$, of a composite system
$\mathcal{C}$, whose Hamiltonian, $\hat H_C$, can be written as:

\begin{equation}
\hat H_S + \hat H_E + \hat H_{SE} = \hat H_C = \hat H_{S'} + \hat
H_{E'} + \hat H_{S'E'}.
\end{equation}

Due to the interaction $\hat H_{SE}$, the entanglement in $S+E$ is
expected to be of the general form $\sum_i \alpha_i \vert \psi_i
\rangle_S \vert \chi_i \rangle_E$--e.g. an instantaneous Schmidt
form of state $\vert \Phi\rangle_C$ of the composite system
$\mathcal{C}$. On the other side, for the separable interaction
[15] $\hat H_{S'E'}$, one obtains for the {\it same state} $\vert
\Phi\rangle_C$ (in the {\it same instant} of time) another form
$\sum_j \beta_j \vert \phi_j \rangle_{S'} \vert
\varphi_j\rangle_{E'}$, still with the equality:

 \begin{equation}
 \sum_i \alpha_i \vert \psi_i \rangle_S \vert \chi_i \rangle_E = \vert \Phi\rangle_C =
 \sum_j \beta_j \vert \phi_j \rangle_{S'} \vert \varphi_j\rangle_{E'}.
 \end{equation}

Independently of their physical contents, the equalities like eq.
(8) emphasize a challenging mathematical task. In the formal
mathematical context, deriving the lhs (rhs) of eq. (8) from the
rhs (lhs) of eq. (8) is an open issue weakly investigated so far.
For certain simple models, one can show [21, 22] that a state
given in a separable form for a decomposition $S + E$ bears
quantum entanglement regarding another decomposition $S' + E'$;
the decompositions being related by certain LCTs. As our dynamical
arguments can hardly cover these methodological gaps for obtaining
the exact (kinematical) forms of the $\mathcal{C}$'s state for the
different decompositions, we do not report any progress in this
regard.

 In the respective position-representations of eq. (6),  eq. (8) reads (up to a constant):

 \begin{equation}
 \sum_i \alpha_i \psi_i(x_{Sm}) \chi(X_{En}) = \sum_j \beta_j \phi_j(\xi_{S'p}) \varphi_j ({\Xi_{E'q}}).
 \end{equation}

\noindent Of course, the presence of  entanglement is not
sufficient for the occurrence of decoherence. But decoherence
requires entanglement.
 Interestingly enough, the occurrence of decoherence for the QBM
 model of Section 2 is
quite independent of the initial correlations in the composite
system [3, 5] (and the references therein).

Now, one may pose the following question: Does the parallel
decoherence apply to a general system? In answering this question,
we emphasize: The above analysis bears some subtlety as the global
LCTs can completely change the character of the model of the
composite system. Then, in general, the task of theoretically
predicting decoherence may be challenging.

To see this, we refer to the simple yet paradigmatic models of the
hydrogen atom, and of the QBM model of Section 2. Let us first
emphasize that the kinetic terms are of the same form for every
subsystem. However, the external fields for the constituent
subsystems as well as their mutual interactions nontrivially
change. For the hydrogen atom, as it is well-known, the (Coulomb)
interaction present for the $e+p$ decomposition disappears in the
$CM+R$ decomposition--it becomes the external (Coulomb) field for
the "relative particle" ($R$). Regarding the QBM model (Section
2), both the external fields as well as the interaction for the
subsystems can change, relative to the original decomposition.
Both, $CM$ and $R$ are "placed" in the quadratic external
potentials. If such potentials are present in the original model,
then the characteristic frequencies are changed. The interaction
in $CM+R$ decomposition is formally the same as for $S+E$
decomposition, yet with the different strength. These examples
illustrate the following general rule: all (but the kinetic) terms
of the Hamiltonian for a decomposition can contribute to all (but
the kinetic) the terms of the Hamiltonian for an alternative
decomposition.

Of course, the changes in the form of the interaction and in its
strength [15] provide the different backgrounds for the possible
occurrence of decoherence for the different global structures. The
change in the character and in the strength of interaction can
give rise to a change in the approximations/physical-assumptions
for the alternative decomposition(s). E.g., if the "weak coupling"
and/or the "rotating wave" (the "secular") approximations [5] are
valid for the original decomposition, this need not be the case
for an alternative global decomposition of the composite system;
similarly, the "spectral density" [5] can change for the different
decompositions. On the other hand, even if the "original
environment" is in thermal equilibrium, the "new environment" need
not be even stationary. Finally, the {\it global} LCT can change
the character of the quantum state of the composite system): a
separable state for one decomposition typically obtains entangled
form for some alternative decomposition [17-24]. Then, a
completely-positive dynamics for one decomposition ($S+E$) can
become non-completely positive dynamics for another structure
($S'+E'$).

Therefore, we answer the above-posed question as follows:
investigating the parallel occurrence of decoherence is a {\it
new} challenging task in the foundations of the decoherence
theory. Bearing in mind the details that may determine the open
system's dynamics, the occurrence of decoherence for the
alternative structures, in general, can not be guaranteed. Rather,
it's a matter of details that should be separately considered for
a class of the similar models of open systems and their
environments.

\bigskip

{\bf 4. Discussion}

\bigskip

\noindent Crucial for our results and observations are  {\it
global}  and mutually {\it irreducible} structures (partitions
into subsystems) of an isolated composite system. While
decoherence regarding the local and/or mutually reducible
structures may bear some subtlety yet to be discovered, the {\it
parallel decoherence} as introduced in this article is
characteristic for the mutually global and irreducible  structures
The "parallel decoherence" means {\it simultaneous}, mutually {\it
non-intersecting} (dynamically independent) unfolding of
decoherence for the different structures. So, at variance with the
standard view, {\it a composite system may host the different,
mutually independent global quasi-classical structures}. The
concept of the (global) quasi-classical structure [4, 5, 10, 11]
{\it is relative}. We believe that this relativity of "structure"
may enrich the classical concept of complexity [25] and may be of
interest for a number of the disciplines. E.g., bearing in mind
eq. (6), a comparison between the two open systems, $S$ and $CM$,
is vague, not only on the intuitive ground. Actually, the two sets
of states appearing in eq. (9), $\{\psi_i(x_{Sm})\}$ and
$\{\phi_j(\xi_{S'p})\}$, do not belong to the same probability
space, neither e.g. $\int \vert \phi_j (\xi_{S'p})\vert^2 \Pi_n
dx_{En}$ can be interpreted as the probability density for $S$.
Consequently, the  complexity of the two semiclassical structures
may be different both in the classical [25] as well as in the
quantum-mechanical context [26] (and references therein). To this
end, the work is in progress and the results will be presented
elsewhere.

Regarding our QBM model of Section 2, one may pose the following
question: Are there only two possible decompositions which give
rise to pointer states? Section 2 and Appendix 1 implicitly answer
this question. Actually, formally every linear canonical
transformation {\it not} involving the momentums preserves the
linear position-position coupling and the physical kind of the
environment that are essential for our finding. Unless the
self-Hamiltonians for the new subsystems appear non-realistic, one
obtains the same conclusion. So, for such type of LCTs, while the
details may be different, we can answer the question: there is
more than two decompositions supporting the QBM effect.

The decoherence-based structure-analysis is not restricted to the
"massive" quantum particles. The LCT can be defined for both, the
"massive particles" (e.g. atoms) interacting with a quantum field
(e.g. the electromagnetic field [27]) as well as to the
interacting quantum fields. While the details  can be different,
as long as the reduced dynamics is Markovian and the coupling is
linear in the transformations-related observables, our finding of
the parallel occurrence of decoherence may   be expected to be
valid. The details in this regard will be presented elsewhere.

 Bearing in mind the global structures, our main result can
be described as given in Introduction: if decoherence establishes
"the appearance of {\it a} classical world" [4] (e.g. $S+E$), our
findings suggest "the appearance of {\it the} classical world{\it
s}" (e.g. $S+E$ {\it and} $CM+R$). As a corollary of the standard
decoherence theory [4, 5, 10, 11], the parallel occurrence of
decoherence opens the following question: This parallel
decoherence implies that the emergent classical world is not
unique, which does not seem supported by our general observations.
Does then the parallel decoherence suggest a further selection
process is required?

The answer to this question is essentially interpretational.
Detailed analysis and arguments in this regard require some space.
Here we just emphasize: If for some interpretational reasons only
one structure is expected to be physically realistic, a selection
rule is needed as emphasized in Zanardi's [18] "Without further
physical assumption, no partition has an ontologically superior
status with respect to any other.", likewise by Halliwell (chapter
3 in Ref. [28]), "However, for many macroscopic systems, and in
particular for the universe as a whole, there may be no natural
split into distinguished subsystems and the rest, and another way
of identifying the naturally decoherent variables is required.".
Further details can be found in Ref. [29].

\bigskip

{\bf 5. Conclusion}

\bigskip

\noindent We launch a search for the occurrence of decoherence
regarding the different global and mutually irreducible
decompositions into subsystems of an isolated  composite quantum
system. For the QBM-similar models, we obtain the occurrence of
decoherence for an alternative decomposition of the composite
system "(open) system plus environment". Physically, this finding
provides us with the observation of the parallel (simultaneous)
occurrence of decoherence thus exhibiting relativity of the basic
physical concept of "(global) semiclassical structure". Being at
variance with the standard view to "physical structure", our
findings open both some conceptual issues yet to be explored and
possibly a new route in describing the composite physical systems.

\bigskip

{\bf Appendix 1}

\bigskip

The two models, eq. (4) and eq. (5), differ in: (a) the values of
the model parameters such as the masses and the characteristic
frequencies, (b) (non)appearance of the external fields for the
respective open systems as well as (c) appearance of the internal
interaction for the new environment, $\hat V_R$, which makes the
model eq. (5) non-linear. Finally, as the two models are mutually
related by the variables (the LCTs) transformations, the state for
the composite system $\mathcal{C}$, which is assumed to be
separable regarding the original structure $S+E$ is now (d)
expected to be of the nonseparable form for $CM+R$ structure.

While the points (a) and (b) are particularly trivial [2-5], the
points (c) and (d) should be carefully examined in the context of
the occurrence of decoherence. We should first emphasize that the
point (c) can be straightforwardly managed by the proper linear
transformations, $\hat \rho_{R\alpha} = \sum_l \lambda_{l\alpha}
\hat Q_{Rl}$, introducing the normal coordinates $\hat Q_{Rl}$.
Then the environmental Hamiltonian $\hat H_{R}$ is linearized,
i.e. obtains the form $\hat H_R = \sum_l (\hat P_{Rl}^2/2 +
\omega^2_l \hat Q_{Rl}^2/2)$ that removes the 'nonlinear term' of
the form of $\hat V_R$ in eq. (5). As this is the linear
transformation referring only to the environment $R$, the coupling
term $\hat X_{CM} \sum_{\alpha} \sigma_{\alpha} \hat
\rho_{R\alpha}$ in eq. (5) acquires another {\it linear} form
$\hat X_{CM} \sum_l \lambda_l \hat Q_{Rl}$ that keeps the form of
the "spectral density" [5].

Regarding the point (d): the introduction of the normal
coordinates for $R$ introduces further change in the $R$'s reduced
state; e.g. a separable state becomes non-separable (entanglement
relativity [17-24]). But this does not constitute any problem here
as the environment is traced out and the tracing-out operation is
basis-independent.

Therefore, the linearization of the Hamiltonian $\hat H_R$ in eq.
(5) gives the form of the Hamiltonian $\hat H$ of the fully
isomorphic form as the 'original' form eq. (4). Then the standard
results of the QBM theory [2-5] directly provide the following
conclusion: likewise the open system $S$ eq. (4), the open system
$CM$ eq. (5) is subject to the QBM-effect, i.e. to the occurrence
of decoherence, which distinguishes the related Gaussian states as
the preferred states for the system $CM$. The "decoherence
function", $\Gamma (t)$,  is of the same form for  both
structures, cf. e.g. eq. (4.226) in Ref. [5]:

\begin{equation}
\Gamma (t) \approx - \vert \alpha - \beta \vert^2 \Lambda (t)/2,
\end{equation}

\noindent for a pair of two "coherent states", $\vert
\alpha\rangle$ and $\vert \beta \rangle$. In the long time limit,
$\Gamma (t)$ is dominated by the overlap, $\vert \alpha - \beta
\vert^2$ [5]. Nevertheless, the "long time limit" may refer to
even mutually incomparable time intervals for the two structures.
So, one may wonder if there is significantly different
"decoherence times" for the two structures, i.e. for the two
decoherence processes. To see this is not the case, some care is
needed.

Actually, even if for an instant of time $t$ for which the ratio
of $\Lambda (t)$ and $\Lambda'(t)$ for the two structures is far
from unity, there is always the possibility also to change the
first factor in eq. (10) in order to obtain $\Gamma (t)$ to be of
the same order for both structures: $\vert \alpha - \beta \vert^2
\Lambda (t) \sim \vert \alpha' - \beta' \vert^2 \Lambda' (t)$.
Physically, it means that, in such cases, the same "decoherence
time" refers to the different Gaussian states for the two
structures. So, the decoherence times are of the {\it same order
of magnitude} for the two structures, yet in general for the
different pairs of the respective Gaussian states.

In effect: the unique unitary dynamics for the composite system
$\mathcal{C}$--generated by the {\it unique} system-Hamiltonian
for the {\it unique} initial state--hides the two, mutually
independent, irreducible and simultaneously occurring decoherence
processes for the two, mutually irreducible open systems, $S$ and
$CM$, that are the subsystems of the mutually irreducible global
structures of the composite system $\mathcal{C}$.

\bigskip

{\bf Acknowledgments}. The work on this paper is financially
supported by Ministry of Science Serbia grant no 171028.

\bigskip

{\bf REFERENCES}

\bigskip

[1] Zurek W H 1998 {\it Philos. Trans. R. Soc. London, Ser.} A
{\bf 356} 1793

[2] Caldeira A O and Leggett A J 1985 {\it Phys. Rev.} A {\bf 31}
1059

[3] Romero L D, Paz J P 1997  {\it Phys. Rev.} A {\bf 55} 4070

[4] Giulini D, Joos E, Kiefer C, Kupsch J, Stamatescu I-O, and Zeh
H D 1996 {\it Decoherence and the Appearance of a Classical World
in Quantum Theory} (Berlin: Springer)

[5] Breuer H P and  Petruccione F 2002 {\it The Theory of Open
Quantum Systems} (Oxford: Clarendon Press)

[6] Gell-Mann M, Hartle J B 1993  {\it Physical Review D} {\bf 47}
3345

[7] Lombardo F C, Villar P I 2006 {\it Int. J. Mod. Phys.} {\bf
B20} 2952

[8] Flores J C 1998 {\it J. Phys. A: Math. Gen.} {\bf 31} 8629

[9] Palma G M, Suominen K-A, Ekert A 1996 {\it Proc. R. Soc.
Lond.} A {\bf 452} 567

[10] Zurek W H 2003 Rev. Mod. Phys. {\bf 75} 715

[11] Schlosshauer M 2004 {\it Rev. Mod. Phys.} {\bf 76} 1267

[12] Banerjee S and Kupsch J 2005 {\it J. Phys. A: Math. Gen.}
{\bf 38} 5237

[13] Schulman L S 1998 {\it Phys. Rev.} A {\bf 57} 840

[14] Schulman L S 2004 {\it Phys. Rev. Lett.} {\bf 92} 210404

[15] Dugi\' c M 1997 {\it Phys. Scr.} {\bf 56} 560

[16] Zeh H D 2005 {\it Seminaire Poincar\' e} {\bf 2} 1

[17] Dugi\' c  M 1999 arXiv:quant-ph/9903037v1

[18] Zanardi P 2001 {\it Phys. Rev. Lett.} {\bf 87} 077901

[19] Dugi\' c M, Jekni\' c J 2006 {\it Int. J. Theor. Phys.} {\bf
45}, 2215

[20] Dugi\' c M, J. Jekni\' c-Dugi\' c J 2008  {\it Int. J. Theor.
Phys.} {\bf 47} 805

[21] De la Torre A C {\it et al} 2010 {\it Europ. J. Phys.} {\bf
31} 325

[22] Harshman N L, Wickramasekara S 2007 {\it Phys. Rev. Lett.}
{\bf 98} 080406

[23] Terra Cunha M O, Dunningham J A,   Vedral V 2007 {\it Proc.
R. Soc.} A  {\bf 463} 2277

[24] Ciancio E, Giorda P, Zanardi P 2006 {\it Phys. Lett.} A {\bf
354} 274

[25] Auyang S Y 1998  {\it Foundations of complex-system theories
in economics, evolutionary biology, and statistical physics},
(Cambridge: Cambridge University Press)

[26] Mora C, Briegel H and Kupsch J 2007 {\it Int. J. Quant. Inf.}
{\bf 5} 729

[27] Bellomo B, Campogno G, Petruccione F 2005 {\it J. Phys. A:
Math. Gen.} {\bf 38} 10203

[28] Saunders S, Barrett J, Kent A, Wallace D (eds.) 2010 {\it
Many Worlds? Everett, Quantum Theory, and Reality} (Oxford: Oxford
University Press)

[29] J. Jekni\' c-Dugi\' c, M. Dugi\' c, A. Francom,
arXiv:1109.6424v1 [quant-ph]

\end{document}